\documentstyle[epsfig]{mn}
\newif\ifAMStwofonts

\newcommand{\n}{\tilde{n}}
\newcommand{\I}{\tilde{I}}

\newcommand{\tl}[1]{\tilde{#1}}
\newcommand{\mt}[1]{\mbox{$\mathbfss{#1}$}}
\newcommand{\colvec}[4]{\left(\begin{array}{c}#1\\#2\\#3\\#4\end{array}\right)}
\newcommand{\pdv}[2]{\frac{\partial #1}{\partial #2}}
\newcommand{\pddv}[2]{\frac{\partial^2 #1}{\partial #2^2}}

\ifoldfss
  \ifCUPmtlplainloaded \else
    \NewTextAlphabet{textbfit} {cmbxti10} {}
    \NewTextAlphabet{textbfss} {cmssbx10} {}
    \NewMathAlphabet{mathbfit} {cmbxti10} {} 
    \NewMathAlphabet{mathbfss} {cmssbx10} {} 
  \fi
  \ifAMStwofonts
    \ifCUPmtlplainloaded \else
      \NewSymbolFont{upmath} {eurm10}
      \NewSymbolFont{AMSa} {msam10}
      \NewMathSymbol{\upi}     {0}{upmath}{19}
      \NewMathSymbol{\umu}     {0}{upmath}{16}
      \NewMathSymbol{\upartial}{0}{upmath}{40}
      \NewMathSymbol{\leqslant}{3}{AMSa}{36}
      \NewMathSymbol{\geqslant}{3}{AMSa}{3E}

    \fi
  \fi
\fi 

\ifnfssone
  \addtoversion{normal}{\mathit}{cmr}{m}{it}
  \addtoversion{bold}{\mathit}{cmr}{bx}{it}
  \newmathalphabet{\mathbfit} 
  \addtoversion{normal}{\mathbfit}{cmr}{bx}{it}
  \addtoversion{bold}{\mathbfit}{cmr}{bx}{it}
  \newmathalphabet{\mathbfss} 
  \addtoversion{normal}{\mathbfss}{cmss}{bx}{n}
  \addtoversion{bold}{\mathbfss}{cmss}{bx}{n}
  \ifAMStwofonts
    \ifCUPmtlplainloaded \else
      \UseAMStwoboldmath
      \makeatletter
      \new@mathgroup\upmath@group
      \define@mathgroup\mv@normal\upmath@group{eur}{m}{n}
      \define@mathgroup\mv@bold\upmath@group{eur}{b}{n}
      \edef\UPM{\hexnumber\upmath@group}
      \new@mathgroup\amsa@group
      \define@mathgroup\mv@normal\amsa@group{msa}{m}{n}
      \define@mathgroup\mv@bold\amsa@group{msa}{m}{n}
      \edef\AMSa{\hexnumber\amsa@group}
      \makeatother
      \mathchardef\upi="0\UPM19
      \mathchardef\umu="0\UPM16
      \mathchardef\upartial="0\UPM40
      \mathchardef\leqslant="3\AMSa36
      \mathchardef\geqslant="3\AMSa3E
    \fi
  \fi
\fi 

\ifnfsstwo
  \DeclareMathAlphabet{\mathbfit}{OT1}{cmr}{bx}{it}
  \SetMathAlphabet\mathbfit{bold}{OT1}{cmr}{bx}{it}
  \DeclareMathAlphabet{\mathbfss}{OT1}{cmss}{bx}{n}
  \SetMathAlphabet\mathbfss{bold}{OT1}{cmss}{bx}{n}
  \ifAMStwofonts
    \ifCUPmtlplainloaded \else
      \DeclareSymbolFont{UPM}{U}{eur}{m}{n}
      \SetSymbolFont{UPM}{bold}{U}{eur}{b}{n}
      \DeclareSymbolFont{AMSa}{U}{msa}{m}{n}
      \DeclareMathSymbol{\upi}{0}{UPM}{"19}
      \DeclareMathSymbol{\umu}{0}{UPM}{"16}
      \DeclareMathSymbol{\upartial}{0}{UPM}{"40}
      \DeclareMathSymbol{\leqslant}{3}{AMSa}{"36}
      \DeclareMathSymbol{\geqslant}{3}{AMSa}{"3E}
    \fi
  \fi
\fi 

\ifCUPmtlplainloaded \else
  \ifAMStwofonts \else 
    \def\upi{\pi}
    \def\umu{\mu}
    \def\upartial{\partial}
  \fi
\fi

\title[Compton scattering of polarized radiation] {The radiative
  transfer equations for Compton scattering of polarized low frequency
  radiation on a hot electron gas} 

\author[Frode K. Hansen and Per B.  Lilje] {{Frode K.
    Hansen\thanks{Present address: Theoretical Astrophysics Center,
      Juliane Maries vej 30, DK-2100 Copenhagen \O,
      Denmark}\thanks{E-mail: fhansen@tac.dk}} and {Per B.
    Lilje\thanks{E-mail: per.lilje@astro.uio.no}}\\ Institute of
  Theoretical Astrophysics, University of Oslo, P. O. Box 1029
  Blindern, N-0315 Oslo, Norway}
\pagerange{\pageref{firstpage}--\pageref{lastpage}} \pubyear{1999}

\begin{document}

\label{firstpage}

\maketitle

\begin{abstract}
  We deduce the equations that describe how polarized radiation is
  Comptonized by a hot electron gas. Low frequencies are considered,
  and the equations are expanded to second order in electron
  velocities. Induced scattering terms are included and a Maxwellian
  velocity distribution for the electrons is assumed. The special case
  of an axisymmetric radiation field is also considered, and the
  corresponding radiative transfer equations are found. Our results
  correct errors and misprints in previosly published transfer
  equations. The extension to a moving electron gas is made, and the
  radiative transfer equations are deduced to second order in gas
  velocity. We use the equations to study polarization in the
  Sunyaev-Zeldovich effect.
\end{abstract}
\section{introduction}
A study of the spectral evolution of a radiation field caused by
Compton scattering on an electron gas was first published by
Kompaneets \shortcite{komp}. He obtained the kinetic equation
describing the evolution of an isotropic and homogenous radiation
field in an infinite, homogenous, thermal and nearly nonrelativistic
electron gas. The Kompaneets equation was later extended to include a
non-homogenous radiation field in a plane parallel atmosphere
\cite{babel}. More recently the extension to arbitrarily high orders
in frequency and temperature has been discussed
\cite{stebbins,challinor}.

Compton scattering of polarized radiation was studied by Chandrasekhar
\shortcite{chandra} in the zero temperature limit. Stark
\shortcite{stark} attempted to extend this work to nonzero
temperatures. He obtained radiative transfer equations for the $I$ and
$Q$ Stokes parameters for low frequency radiation scattered on a hot,
nearly nonrelativistic thermal electron gas, including induced
scattering and assuming an axisymmetric radiation field. The equations
were derived by expanding to first order in $kT/mc^2$ and $h\nu/mc^2$,
where $h\nu$ is photon energy, $kT$ is thermal energy and $mc^2$ is
the electron rest energy.

The fully relativistic equations for all four Stokes parameters were
obtained by Nagirner and Poutanen \shortcite {nagpot} and further
extended to include induced scattering by Nagirner \shortcite
{nagirner}. These equations are very complicated and hard to use in
practice. Therefore, when working with low frequencies and electron
velocities, a simplified form is preferred. Nagirner did also reduce
the fully relativistic equations in this limit, to the same order as
Stark \shortcite{stark}.

However, when the simplified equations of Nagirner
\shortcite{nagirner} are integrated assuming an axisymmetric field,
the result is very different from the equations obtained by Stark
\shortcite{stark}. The two sets of equations are supposed to describe
the same phenomenon, but several terms disagree. To make the situation
even more confusing, the equations in the Russian original paper by
Nagirner \shortcite{nagirner2} are different from those in the English
translation \cite{nagirner}. As we show in this paper, the reason for
the disagreements is that some relativistic corrections to the
rotation angles between different polarization bases were neglected by
Stark. Some small integration errors in both papers also contribute to
the differences.

In this paper we present the correct radiative transfer equations for
polarized radiation in an electron gas to first order in $h\nu/mc^2$
and $kT/mc^2$ [equation (\ref{eq:boltzmann3})]. The equations are
derived from the Boltzmann collisional equation. The special case of
an axisymmetric field is considered [equations (\ref{eq:I}) and
(\ref{eq:Q})]. We then extend the equations to include a moving
electron gas, to second order in gas velocity [equation
(\ref{eq:complete})].

The transfer equations for Compton scattering of low frequency
radiation on a relatively hot electron gas is of high importance in
cosmology. Clusters of galaxies contain a hot electron gas, and the
cosmic background radiation is changed because of Compton scattering
on this gas. This is the so called Sunyaev-Zeldovich effect (SZ
effect) \cite{sz1,rephaeli2}. This effect together with X-ray data of
clusters of galaxies is used e.g. to measure the Hubble constant. The
SZ-effect for a moving electron gas \cite{sz2,rephaeli2,sazsun} can
be used to measure peculiar velocities of clusters. Such measurements
have so far only used the intensity of the cosmic background radiation
(CBR). Upcoming satellite experiments like {\it MAP} and {\it Planck}
will try to measure polarization in the CBR. Therefore the equations
describing the change in the Stokes parameters of the CBR caused by
Compton scattering in clusters [equations (\ref{eq:boltzmann3}) and
(\ref{eq:complete}) in this paper] will be important. We investigate
if there is any temperature or velocity dependent polarization
produced. We also briefly discuss scattering of radiation from
extended radio sources in the cluster on the intracluster gas and
possible polarization produced from this effect.

Here we derive the radiative transfer equations, starting in Section
\ref{sect:restframe} with the Boltzmann collisional equations in the
rest frame of an electron. Using Lorentz transformations, we transform
the equations to the rest frame of the electron gas, and then expand
to second order in electron velocity and first order in $h\nu/mc^2$,
before averaging over electron velocities (Section
\ref{sect:integration}). The final result is presented in equation
(\ref{eq:boltzmann3}). We use different bases for the Stokes vector in
the electron rest frame and in the rest frame of the gas. In Section
\ref{sect:bases} we discuss the transformation between these two
bases. In Section \ref{sect:symmetry}, the equations are integrated in
the case of an axisymmetric field to compare with the equations of
Stark [equations (\ref{eq:I}) and (\ref{eq:Q})]. Then, in Section
\ref{sect:moving}, equation (\ref{eq:boltzmann3}) is extended to a
moving electron gas. We derive the radiative transfer equations to
second order in gas velocity [equation (\ref{eq:complete})]. Finally,
applications of the formalism deduced in the previous sections to the
SZ effect is discussed in Section \ref{sect:SZ}. In the conclusions,
we discuss the flaws in the previously published results.  Throughout
this paper, the relativistic quantum system of units, in which the
electron mass, speed of light and Planck's constant are given by
$m=c=\hbar=1$, is used.

\section{The Boltzmann collisional equation for the Stokes parameters}
\label{sect:restframe}
\begin{figure}
\begin{center}
\leavevmode
\epsfig {file=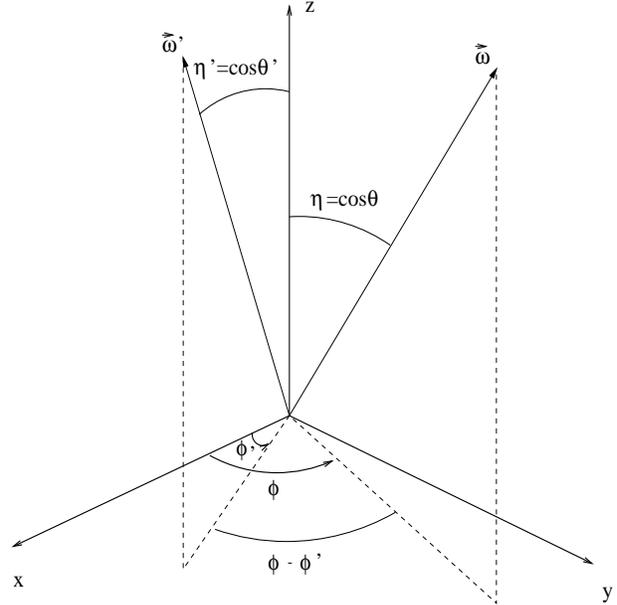,height=8cm,width=8cm}
\caption{The incoming and outgoing photons in the xyz-coordinate system.}
\label{fig:xyzco}
\end{center}
\end{figure}

In this section, we study the evolution of a beam of polarized
radiation in the `system frame', in which the mean momentum of the
electrons in the gas is zero. In this frame we introduce a fixed
coordinate system $(x,y,z)$ with arbitrary directions of the axes. A
photon beam in the direction $(\theta,\phi)$ (see Fig.
\ref{fig:xyzco}) is described by the `number' Stokes vector
\[
\n(\nu,\theta,\phi,t,\mathbf{x})=\colvec{n}{n_Q}{n_U}{n_V},
\] 
where $\nu$ is the frequency, $\mathbf{x}$ is position in space and
$t$ is time. The number Stokes parameters $n, n_Q, n_U$ and $n_V$ are
related to the usual Stokes parameters by
\[
\tl{I}=\colvec{I}{Q}{U}{V}=4\pi\nu^3 \n.
\]

\begin{figure}
\begin{center}
\leavevmode
\epsfig {file=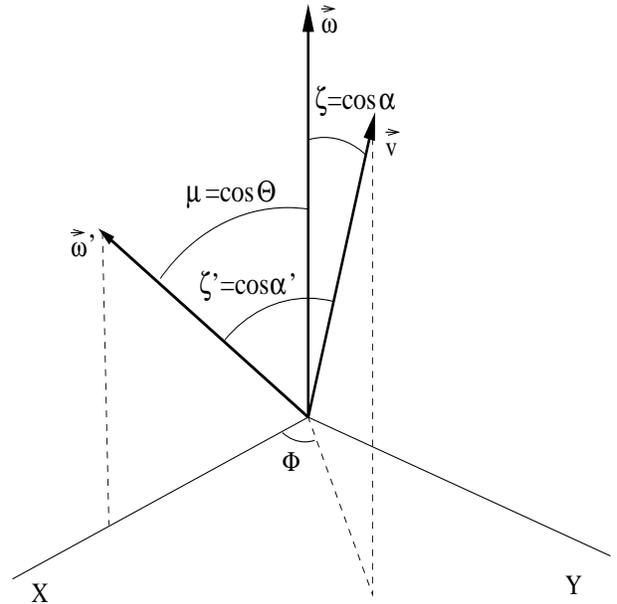,height=8cm,width=8cm}
\caption{The incoming photon beam and the electron velocity in the
  coordinate system where the z-axis is along the outgoing photon
  direction.} 
\label{fig:photonco}
\end{center}
\end{figure}

First we study the photon beam in the rest frame of an electron moving
with velocity $\mathbf{v}$. All quantites taken in this frame are
denoted by subscript `e'. The number Stokes parameters $n, n_Q, n_U,
n_V$ are relativistically invariant, but frequencies and angles
transform according to the usual Lorentz transformations,

\begin{eqnarray}
\label{eq:lorentz}
&\nu=\gamma\nu_e(1+v\cos{\alpha_e}), &
\cos{\alpha}=\frac{\cos{\alpha_e}+v}{1+v\cos{\alpha_e}},\nonumber\\ 
\lefteqn{d\Omega_e=\frac{d\Omega}{\gamma^2(1-v\cos{\alpha})^2},}
\nonumber\\  
\lefteqn{\cos{\Theta_e}=1-\frac{(1-\cos{\Theta})}{\gamma^2(1-v
    \cos{\alpha})(1-v\cos{\alpha'})},}\nonumber\\  
\lefteqn{\pdv{}{t_e}+\bomega_e\cdot\nabla_e=
  \frac{1}{\gamma(1-v\cos{\alpha})}\left(\pdv{}{t}+
  \bomega\cdot\nabla\right),}\nonumber\\   
&\gamma=(1-v^2)^{-1/2},&
\end{eqnarray}
where $(\alpha,\Phi)$ are the polar and azimuthal angles between
$\mathbf{v}$ and the photon [see Fig. \ref{fig:photonco}] and
$d\Omega$ is an infinitesimal solid angle. The direction unit vector
$\bomega$ gives the direction of the photon beam and $\Theta$ is the
angle between two photons which have angles $\alpha$ and $\alpha'$
with the electron velocity $\mathbf{v}$.  The Boltzmann collisional
equation describes the change of
$\n_e(\nu_e,\theta_e,\phi_e,t_e,\mathbf{x_e})$ caused by scattering on
particles with velocity $v$ \cite{aqan}. In the rest frame the
equation takes the form
\begin{equation}
\label{eq:boltzmann1}
\pdv{\n}{t_e}+\bomega_e\cdot\nabla\!_e\n= \int
(d\n_{e}^{+}-d\n_{e}^{-}) F_e(v^2,\rho_e), 
\end{equation}
where $\n_{e}^{+}$ and $\n_{e}^{-}$ are the number of photons per
phase space volume scattered into the beam $(\nu_e,\theta_e,\phi_e)$
and out of it, respectively from each scattering. Assuming an
isotropic electron gas, the number of electrons with velocity $v$ is
given by the electron distribution function $F_e(v^2,\rho_e)$, where
$\rho_e$ is the electron density in the rest frame. Next, we assume
that the distribution function can be written as
$F_e(v^2,\rho_e)=\rho_e f_e(v^2)$, which in the system frame is
$F(v^2,\rho)=\rho\gamma^{-1} f(v^2)$. Here $f(v^2)$ is normalised as
$4\pi\int_{0}^{\infty} f(v^2) v^2 dv=1$. Going to the system frame,
using the Lorentz transformations [equation (\ref{eq:lorentz})] above,
we find that
\begin{equation}
\label{eq:boltzmann}
\pdv{\n}{t}+\bomega\cdot\nabla\n= \int(d\n_{e}^{+}- d\n_{e}^{-})
(1-v\cos{\alpha})\rho f(v^2).
\end{equation}

The rate of scattering from one photon beam
$(\nu_e',\theta_e',\phi_e')$ into $(\nu_e,\theta_e,\phi_e)$ is given
by
\[ d\n_{e}^{+}\nu_e^2d\nu_e=(1+\mt{N})\mt{L}_1\mt{R}_e
\mt{L}_2\nu_e'^{\ 2}d\nu_e' \n_e'd\Omega_e',
\] 
where $\n'=\n(\nu',\theta',\phi')$,
$d\Omega_e'=d\cos{\theta_e'}d\phi_e'$ and $\mt{R}_e$ is the scattering
matrix in the rest frame, given by \cite{beret}
\[
\mt{R}_e=\frac{3}{16\pi}\sigma_T \left(\frac{\nu_e}{\nu_e'}\right)^2\left(
\begin{array}{cccc}
R_{11}&R_{12}&0&0\\
R_{12}&R_{22}&0&0\\
0&0&R_{33}&0\\
0&0&0&R_{44}
\end{array}
\right),\\
\]
where
\begin{eqnarray}
\label{eq:rm}
R_{11}=\left(\frac{\nu_e'}{\nu_e}\right)^2 \frac{16\pi}{3\sigma_T} R_e,&
R_{12}=\cos^2{\Theta_e}-1,\\
R_{22}=\cos^2{\Theta_e}+1,&
R_{33}=2\cos{\Theta_e}\nonumber\\
\rm{and\hspace{3cm}}&\nonumber\\
R_{44}=\cos{\Theta_e}\left(\frac{\nu_e'}{\nu_e}+
\frac{\nu_e}{\nu_e'}\right).&\nonumber\\ \nonumber 
\end{eqnarray}

The Klein-Nishina cross section $R_e$ is given by
\begin{equation}
\label{eq:r}
R_e=\frac{3}{16\pi}\sigma_T  \left(\frac{\nu_e}{\nu_e'}\right)^2
\left[\frac{\nu_e}{\nu_e'}+\frac{\nu_e'}{\nu_e}+
\cos^2{\Theta_e}-1\right],
\end{equation}

where $\sigma_T$ is the Thompson cross section. The \mt{L} matrices
transform between different polarization bases as discussed in the
next section.  The more photons there are in the state
$(\nu_e,\theta_e,\phi_e)$, the more likely it is that photons are
scattered into this state. This fact (induced scattering) is included
by the factor $(1+\mt{N})$, where $\mt{N}$ is given by \cite{nagirner}
\[
\mt{N}=\left(
\begin{array}{cccc}
n&n_Q&n_U&n_V\\
n_Q&n&0&0\\
n_U&0&n&0\\
n_V&0&0&n\\
\end{array}
\right). 
\]
The incoming and outgoing frequencies are related by the usual formula
for the Compton frequency shift, 
\[
\nu_e'=\frac{\nu_e}{1-\nu_e(1-\cos{\Theta_e})}.
\]
The rate of scattering out of the beam $(\nu_e,\theta_e,\phi_e)$ is
\cite{nagirner} 
\[
d\n_e^{-}=(R_e' \n_e+\mt{N}\mt{L}_1\mt{R}'_e\mt{L}_2\n_e'')d\Omega_e',
\] 
where $\n''=\n(\nu'',\theta',\phi')$. Note that the spontanous
scattering term (the first term on the right hand side) in this
equation contains the Klein-Nishina cross section and {\bf not} the
scattering matrix. This is because of the isotropy of the medium
\cite{aqan}. The cross section $R_e'$ and the scattering matrix
$\mt{R}_e'$ are given by equation (\ref{eq:r}) and (\ref{eq:rm}),
except that $\nu_e$ and $\nu_e'$ are replaced by $\nu_e''$ and $\nu_e$
respectively. The frequencies $\nu_e''$ and $\nu_e$ are related by the
Compton formula for the scattering $\nu_e\rightarrow\nu_e''$,
\[
\nu_e''=\frac{\nu_e}{1+\nu_e(1-\cos{\Theta_e})}.
\] 
Inserting the expressions for $d\n_e^-$ and $d\n_e^+$ into equation
(\ref{eq:boltzmann}), we find that
\begin{eqnarray}
\label{eq:boltzmann2}
\pdv{\n}{t}+\bomega\cdot\nabla\n=\int\biggl\{\rho(1-v \cos{\alpha})
f(v^2)\biggl[(1+\mt{N})\mt{L}_1\mt{R}_e\mt{L}_2 
\nonumber\\ 
\left(\frac{\nu_e'}{\nu_e}\right)^2\frac{d\nu_e'}{d\nu_e}\n'-R_e'\n-
\mt{N}\mt{L}_1\mt{R}'_e\mt{L}_2\n''\biggr]\frac{d\Omega_e'}{d\Omega'}
\biggr\}d\Omega'. 
\end{eqnarray}
\section{polarization bases}
\label{sect:bases}
So far, the polarization vector basis has not been considered. The
scattering matrix $\mt{R}_e$ describes the scattering of photons
$\n_e$ in a certain polarization basis which we call the `scattering
basis'. This basis is dependent on the photons and electrons involved
in each scattering event \cite{beret}, and is therefore unconvenient
when considering several scatterings. We wish to describe the
polarization relative to the fixed set of axes $(x,y,z)$ in the system
frame. When we use the vector $\n$ in this frame, we use a
polarization basis which we call the `system basis'. In this basis,
the Stokes parameters are measured relative to the meridian planes.
The matrix $\mt{L}_2$ rotates the vector $\n$ from the system basis to
the scattering basis, so that the matrix $\mt{R}_e$ can be applied.
Then $\mt{L}_1$ rotates back to the system basis \cite{chandra}. A
Stokes vector can be rotated from one basis to another with the
rotation matrix \cite{chandra}
\[
\mt{L}(\chi)=\left(
\begin{array}{cccc}
1&0&0&0\\
0&\cos{2\chi}&\sin{2\chi}&0\\
0&-\sin{2\chi}&\cos{2\chi}&0\\
0&0&0&1\\
\end{array}\right),
\]
where $\chi$ is the rotation angle of the polarization basis
vectors. Using the notation of Chandrasekhar, we write 
\[
\mt{L}_1\mt{R}_e\mt{L}_2=\mt{L}(\pi-i_{2e})\mt{R}_e\mt{L}(-i_{1e}),
\] 
where $\pi-i_{2e}$ and $-i_{1e}$ are the rotation angles in the
rest frame of the electron. These angles must be transformed to the
system frame and expanded to second order in electron velocity. We
wish to express $i_{1e}$ and $i_{2e}$ in terms of the zeroth order
rotation angles $i_1$ and $i_2$ used by Chandrasekhar
\shortcite{chandra}. We separate the factors which are dependent on
electron directions (which we call $x$ and $y$) from $i_1$ and $i_2$
which are not. This enables us, as will be shown in the next section,
to collect all dependency on electron directions (which we average
over) in one matrix, making the integration in Section
\ref{sect:integration} easier. The Lorentz transformed rotation angles
are given by \cite{nagpot}:
\begin{eqnarray}
\label{eq:angles}
\cos{i_{1e}}&=&x\cos{i_1}+y\sin{i_1},\nonumber\\
\sin{i_{1e}}&=&x\sin{i_1}-y\cos{i_1}\nonumber\\
\cos{i_{2e}}&=&x\cos{i_2}+y\sin{i_2},\nonumber\\
\sin{i_{2e}}&=&x\sin{i_2}-y\cos{i_2},\nonumber\\
x&=&\frac{1}{\Delta}\left(\gamma(1-v \zeta)\sqrt{\frac{1+\mu}{1-\mu}}-
\gamma v \sqrt{1-\zeta^2}\cos{\Phi}\right),\nonumber\\
y&=&\frac{1}{\Delta}\gamma v \sqrt{1-\zeta^2}\sin{\Phi},\nonumber\\
x^2+y^2&=&1.
\end{eqnarray}
Here the following definitions were used:
\begin{eqnarray}
\label{eq:mu}
\Delta&=&\sqrt{\frac{2\gamma^2(1-v \zeta)(1-v \zeta')}{(1-\mu)}-1},
\nonumber\\ 
&\zeta=\cos{\alpha}&\zeta'=\cos{\alpha'},\nonumber\\
&\eta=\cos{\theta}&\eta'=\cos{\theta'},\nonumber\\
\mu&=&\eta \eta'+\sqrt{1-\eta^2}\sqrt{1-\eta'^2}\cos{(\phi-\phi')},
\end{eqnarray}
where $\alpha'$ is the angle between the photon $(\theta',\phi')$ and
the electron velocity $\mathbf{v}$. The angle between the two photons
is $\Theta$ and $\mu=\cos{\Theta}$. The angles $i_1$ and $i_2$ are the
zeroth order rotation angles given by Chandrasekhar
\shortcite{chandra},
\begin{eqnarray}
\cos{i_1}&=&\frac{\eta-\eta'\mu}{\sqrt{1-\eta'^2}\sqrt{1-\mu^2}},
\nonumber\\ 
\cos{i_2}&=&\frac{\eta'-\eta\mu}{\sqrt{1-\eta^2}\sqrt{1-\mu^2}},
\nonumber\\ 
\sin{i_1}&=&\sqrt{\frac{1-\eta^2}{1-\mu^2}}\sin{(\phi-\phi')},\nonumber\\
\sin{i_2}&=&\sqrt{\frac{1-{\eta'}^2}{1-\mu^2}}\sin{(\phi-\phi')}.
\label{eq:zerothi12}
\end{eqnarray}
\section{integration over electron velocities}
\label{sect:integration}
The next step is to average over alle possible electron directions by
integrating the right side of equation (\ref{eq:boltzmann2}) over
$\zeta$ from $0$ to $1$ and $\Phi$ from $0$ to $2\pi$ and divide by
$2\pi$. Before integrating, all terms are expanded to second order in
$v$ and first order in $\nu$. Expanding to second order in $v$ is
equivalent to first order in $kT/mc^2$ when using a Maxwellian
distribution function for the electrons. We expand $\n'$ and $\n''$
near $\nu\approx\nu'$ and $\nu\approx\nu''$:
\begin{eqnarray*}
\n(\nu',\theta',\phi')&\approx&\n'+\pdv{\n'}{\nu}(\nu'-\nu)+
\frac{1}{2}\pddv{\n'}{\nu}(\nu'-\nu)^2,\\
\n(\nu'',\theta',\phi')&\approx&\n'+\pdv{\n'}{\nu}(\nu''-\nu)+
\frac{1}{2}\pddv{\n'}{\nu}(\nu''-\nu)^2, 
\end{eqnarray*}
where we redefine $\n'=\n(\nu,\theta',\phi')$. 
Using the Lorentz transformation equations (\ref{eq:lorentz}),
$\nu_e''/\nu_e$ and $d\nu_e'/d\nu_e$ are transformed to give 
\begin{eqnarray*}
\frac{\nu_e''}{\nu_e}&=&\frac{(1-v\zeta')}{(1-v\zeta)}
\frac{\gamma(1-v \zeta)}{\gamma(1-v \zeta')+\nu(1-\mu)},\\ 
\frac{d\nu_e'}{d\nu_e}&=&\left(\frac{\nu_e'}{\nu_e}\right)^2=
\frac{\gamma^2(1-v\zeta')^2}{(\gamma(1-v\zeta')-\nu(1-\mu))^2},\\ 
\frac{d\Omega_e'}{d\Omega'}&=&\frac{1}{\gamma^2(1-v\zeta')^2}.
\end{eqnarray*}
In appendix \ref{app:expansions} we have expanded all quantities we
need in equation (\ref{eq:boltzmann2}) to first order in $\nu$ (low
frequencies) and second order in $v$ (equivalent to first order in
temperature). The quantities $u=2xy$ and $w=(2x^2-1)$ are expanded
instead of $x$ and $y$ for reasons which will become obvious.
 
The rotation matrices $\mt{L}_1$ and $\mt{L}_2$ contain cosine and
sine of the angles $2(i_{2e}-\pi)$ and $2(-i_{1e})$. Again we wish to
separate the factors that are dependent on electron directions from
those that are not. We use equations (\ref{eq:angles}) to write
\begin{eqnarray}
\label{eq:factorangles}
\cos{2(\pi-i_{2e})}&=&(2x^2-1)\cos{2(\pi-i_2)}-2 x y \sin{2(\pi-i_2)},
\nonumber\\
\sin{2(\pi-i_{2e})}&=&(2x^2-1)\sin{2(\pi-i_2)}+2 x y \cos{2(\pi-i_2)},
\nonumber\\
\cos{2(-i_{1e})}&=&(2x^2-1)\cos{2(-i_1)}-2 x y \sin{2(-i_1)},\nonumber\\
\sin{2(-i_{1e})}&=&(2x^2-1)\sin{2(-i_1)}+2 x y \cos{2(-i_1)}.
\end{eqnarray}

Using these equations we can write $\mt{L}_1\mt{R}_e\mt{L}_2$ as
\[
\mt{L}_1\mt{R}_e\mt{L}_2=\mt{L}(\pi-i_2)\mt{A}\mt{L}(-i_1),
\]
where \mt{A} is given by,
\begin{eqnarray*}
&&\mt{A}=\frac{3}{16\pi}\sigma_T \left(\frac{\nu_e}{\nu_e'}\right)^2
\times\\ 
&&\left(
\begin{array}{cccc}
R_{11}&R_{12}w&R_{12}u&0\\
R_{12}w&R_{22}w^2-R_{33}u^2&(R_{22}+R_{33})uw&0\\
-R_{12}u&-(R_{22}+R_{33})uw&R_{33}w^2-R_{22}u^2&0\\
0&0&0&R_{44}\\
\end{array}
\right).
\end{eqnarray*}
First of all, one can see that all terms proportional to $\sin{\Phi}$
can be omitted since they will disappear in the averaging over $\Phi$.
The \mt{L} matrices are now independent of electron directions, so
they can be treated as constants in the integration.  Next, we insert
the expanded expressions into equation (\ref{eq:boltzmann2}) and use
the relation (which can be found from geometric considerations)
\begin{equation}
\label{eq:zetam}
\zeta'=\mu \zeta+\sqrt{1-\mu^2}\sqrt{1-\zeta^2}\cos{\Phi}.
\end{equation}
Finally, $Mathematica^{(R)}$ is used to integrate over $\Phi$ and
$\zeta$. Then we average over velocities, using a non relativistic
Maxwellian distribution,
\begin{equation}
\label{eq:maxwell}
f(v^2)=(2\pi kT)^{-3/2}e^{-v^2/2kT},
\end{equation}
where $k$ is the Boltzmann constant, and $T$ is the temperature of the
gas. Remembering that $\langle v\rangle =0$ and $\langle v^2\rangle
=3kT\equiv3\tau$ where  
\[
\langle X\rangle=\frac{\int_{-\infty}^{\infty}Xf(v^2)v^2dv}
{\int_{-\infty}^{\infty}f(v^2)v^2dv}, 
\]
we find that
\begin{eqnarray}
\label{eq:boltzmann3}
&&\pdv{\n}{t}+\bomega\cdot\nabla\n=-\sigma_T\rho(1-2\nu)\n\nonumber\\ 
&&+\sigma_T\rho\frac{3}{16\pi}\int\biggl\{\mt{L}(\pi-i_2) \mt{B}
\mt{L} (-i_1)\biggl[\n'+(1-\mu)(2\nu\n'\nonumber\\ 
&&+(\nu+4
\tau)\nu\left(\pdv{\n'}{\nu}\right)+\tau\nu^2\left(\pddv{\n'}{\nu}
\right))\biggl]\nonumber\\  
&&-2\tau\mt{L}(\pi-i_2)\mt{C}\mt{L}(-i_1)\n'\nonumber\\ 
&&+2(1-\mu)\nu\mt{N}\mt{L}(\pi-i_2)\mt{B}\mt{L}(-i_1)\biggl[2\n'+
\nu\left(\pdv{\n'}{\nu}\right)\biggr]\biggr\}d\Omega',\nonumber\\  
\end{eqnarray}
where the $\mt{B}$ and $\mt{C}$ matrices are defined in appendix 
\ref{app:matrices}. 

These are the radiative transfer equations for all four Stokes
parameters in a rather hot thermal isotropic electron gas (first order
in $kT/mc^2$) for low frequency (first order in $h\nu/mc^2$) radiation
in the rest frame of the electron gas. If the equation for the
I-Stokes parameter is integrated assuming no dependency on angles in
the radiation field $\n$ (isotropic radiation field) we recover the
Kompaneets equation \cite{komp}.
\section{The radiative transfer equation for an axisymmetric field}
\label{sect:symmetry}
Now the field $\n$ is assumed homogenous and symmetric about the
$z$-axis, so there is no $\phi$ dependence. With this symmetry, there
is no $U$-polarization, so only the $I$ and $Q$ components of the
Stokes vector are considered. We also change to the intensity
parameters $(I,Q)$ by means of the relations
\begin{eqnarray*}
\n=\frac{\I}{4\pi\nu^3},&&\pdv{\n}{\nu}=\frac{1}{4\pi\nu^3}
\left(\pdv{\I}{\nu}-\frac{3}{\nu}\I\right),\\ 
\pddv{\n}{\nu}&=&\frac{12}{4\pi\nu^5}\I-\frac{6}{4\pi\nu^4}\pdv{\I}{\nu}+
\frac{1}{4\pi\nu^3}\pddv{\I}{\nu}.\\ 
\end{eqnarray*}
Making the substitution $z=\cos{(\phi-\phi')}$, we define
\begin{eqnarray*}
\mt{D}&\equiv&\frac{2}{\pi}\int_{-1}^{1}\frac{dz}{\sqrt{1-z^2}}
\left[\mt{L}(\pi-i_2)\mt{B}\mt{L}(-i_1)\right]_{ul},\\ 
\mt{E}&\equiv&\frac{2}{\pi}\int_{-1}^{1}\frac{dz}{\sqrt{1-z^2}}\mu
\left[\mt{L}(\pi-i_2)\mt{B}\mt{L}(-i_1)\right]_{ul},\\ 
\mt{F}&\equiv&\frac{2}{\pi}\int_{-1}^{1}\frac{dz}{\sqrt{1-z^2}}
\left[\mt{L}(\pi-i_2)\mt{C}\mt{L}(-i_1)\right]_{ul},\\ 
\end{eqnarray*}
where $[\mt{M}]_{ul}$ means the upper left minor of the matrix \mt{M}.
Now, using $\mu=\eta \eta'+\sqrt{1-\eta^2}\sqrt{1-\eta'^2} z $ to
integrate over $z$, we find
\begin{eqnarray*}
\mt{D}=\left(
\begin{array}{cc}
3-\eta^2-\eta'^2(1-3\eta^2)&
(3\eta^2-1)(\eta'^2-1)\\
(\eta^2-1)(3\eta'^2-1)&
3(1-\eta^2)(1-\eta'^2)\\
\end{array}
\right),\\
\mt{E}=\eta\eta'\left(
\begin{array}{cc}
5-3\eta^2-\eta'^2(3-5\eta^2)&
(5\eta^2-3)(\eta'^2-1)\\
(\eta^2-1)(5\eta'^2-3)&
5(1-\eta^2)(1-\eta'^2)
\end{array}
\right)\\
\end{eqnarray*}
and
\begin{eqnarray*}
F_{11}&=&1-3\eta'^2+3\eta^2(3\eta'^2-1)+\\
&&2\eta^3\eta'(3-5\eta'^2)+2\eta\eta'(3\eta'^2-1),\\
F_{12}&=&-2(\eta'^2-1)(1-3\eta^2-3\eta\eta'+5\eta^3\eta'),\\
F_{21}&=&-2(\eta^2-1)(1-3\eta\eta'-3\eta'^2+5\eta'^3\eta),\\
F_{22}&=&(\eta^2-1)(\eta'^2-1)(3-10\eta\eta').\\
\end{eqnarray*}
Finally we define (using the notation of Stark 1981)
\[
X_{n}\equiv\int_{-1}^{1}X\eta'^n d\eta',
\]
and 
\[
F\equiv\nu\left(-1+\nu\pdv{\ }{\nu}\right)+\tau \left(\nu^2\pddv{\
  }{\nu}-2\nu\pdv{\ }{\nu}\right), 
\] 
where $X=(I,Q)$ and $n=(0,1,2,3)$. With this notation, the
radiative transfer equations for $I$ and $Q$ take the form
\begin{eqnarray}
\label{eq:I}
\pdv{I}{s}&=&-I(1-2\nu)+\frac{3}{16}\bigl\{(3-\eta^2)I_0+
(1-3\eta^2)(Q_0-I_2\nonumber\\ 
&&-Q_2)+F[G_I]+2\tau[(3\eta^2-1)(I_0-2\eta I_1-3I_2)\nonumber\\
&&+2(5\eta^2-3)\eta(I_3+Q_3-Q_1)+2(3\eta^2-1)(Q_0\nonumber\\
&&-Q_2)]+\frac{\nu}{\nu^3}\bigl[I(-1+\nu\pdv{\ }{\nu})G_I+
Q(-1+\nu\pdv{\ }{\nu})G_Q\bigr]\bigr\},\nonumber\\ 
\end{eqnarray}

\begin{eqnarray}
\label{eq:Q}
\pdv{Q}{s}&=&-Q(1-2\nu)+\frac{3}{16}\bigl\{(1-\eta^2)
(I_0+3Q_0-3Q_2-3I_2)\nonumber\\ 
&&+F[G_Q]+2\tau[(1-\eta^2)(-2I_0+6(\eta I_1+I_2)\nonumber\\
&&-3(Q_0-Q_2)+10\eta(Q_1-I_3-Q_3))]\nonumber\\
&&+\frac{\nu}{\nu^3}\bigl[Q(-1+\nu\pdv{\ }{\nu})G_I+ I(-1+\nu\pdv{\
  }{\nu})G_Q\bigr]\bigr\},\nonumber\\ 
\end{eqnarray}
where $s=\sigma_T\rho t$, and 
\begin{eqnarray*}
G_I&\equiv&(3-\eta^2)I_0+(1-3\eta^2)(Q_0-I_2-Q_2)\\
&&-\eta((5-3\eta^2)I_1+(3-5\eta^2)(Q_1-I_3-Q_3)),\\
G_Q&\equiv&(1-\eta^2)(I_0+3Q_0-3Q_2-3I_2\\
&&-\eta(3I_1+5Q_1-5Q_3-5I_3)).\\
\end{eqnarray*}
Equations (\ref{eq:I}) and (\ref{eq:Q}) are in Section
\ref{sect:concl} compared with the results of Stark \shortcite{stark}.
\section{The radiative transfer equations for Compton scattering of 
polarized low frequency radiation on a moving electron gas}
\label{sect:moving}
In Sections \ref{sect:bases} and \ref{sect:integration} we transformed
the Boltzmann collisional equations from the rest frame of the
electron [equation (\ref{eq:boltzmann})] to the rest frame of the
electron gas. Now we obtain the equations in a frame of reference
where the electron gas is moving. In this frame the electrons have a
common nonzero mean velocity. Therefore, we now have to make a
transformation from the rest frame of the electron to a frame where
the electron has a velocity which is a sum of the thermal velocity
(the velocity in the rest frame of the electron gas) and the velocity
of the gas.

We expand to second order in thermal velocity ($v$) and electron gas
velocity ($v_g$). To this order, relativistic addition of velocities
reduces to $\mathbf{v_t}=\mathbf{v}+\mathbf{v_g}$ where $\mathbf{v_t}$
is the total electron velocity observed from the moving frame,
$\mathbf{v}$ is the velocity in the rest frame of the gas (thermal
velocity) and $\mathbf{v_g}$ is the velocity of the gas observed in
the moving frame. Using spherical coordinates, we add each of the
components to get
\begin{eqnarray}
\label{eq:addvel}
v_t\sin{\alpha_t}\cos{\Phi_t}&=&v\sin{\alpha}\cos{\Phi}+v_g\sin{\alpha_g}\cos{\Phi_g},\nonumber\\
v_t\sin{\alpha_t}\sin{\Phi_t}&=&v\sin{\alpha}\sin{\Phi}+v_g\sin{\alpha_g}\sin{\Phi_g},\nonumber\\ 
v_t\cos{\alpha_t}&=&v\cos{\alpha}+v_g\cos{\alpha_g},
\end{eqnarray}
where $v_t=|\mathbf{v_t}|$, $v_g=|\mathbf{v_g}|$. Here
$(\alpha_t,\Phi_t)$ are the angles between the total velocity vector
and the photon, $(\alpha,\Phi)$ are the angles between the electron
velocity and the photon in the rest frame of the gas and
$(\alpha_g,\Phi_g)$ are the angles between the gas velocity and the
photon.

We follow the same steps that we used to obtain the equations in the
rest frame of the electron gas. The difference is that we now
transform to a frame where the electron velocity is
$\mathbf{v_t}=\mathbf{v}+\mathbf{v_g}$ instad of $\mathbf{v}$. So, in
equation (\ref{eq:boltzmann2}) and in the expanded quantities
[equation (\ref{eq:expanded})], we have to make the following
replacements [using equations (\ref{eq:addvel})]:
\begin{eqnarray*}
v\sqrt{1-\zeta^2}\cos{\Phi}&\longrightarrow& v\sqrt{1-\zeta^2}\cos{\Phi}+v_g\sqrt{1-\zeta_g^2}\cos{\Phi_g},\\
v\sqrt{1-\zeta^2}\sin{\Phi}&\longrightarrow& v\sqrt{1-\zeta^2}\sin{\Phi}+v_g\sqrt{1-\zeta_g^2}\sin{\Phi_g}\\
v\zeta&\longrightarrow& v\zeta+v_g\zeta_g,
\end{eqnarray*}
where $\zeta_g=\cos{\alpha_g}$. Making these replacements, we write
the right hand side of equation (\ref{eq:boltzmann2}) as
\begin{eqnarray*}
&&F_{0}+F_{1}(\zeta_g,\Phi_g)v_g+F_{1}(\zeta,\Phi)v+F_{2}(\zeta_g,\Phi_g)v_g^2\\ 
&&+F_{2}(\zeta,\Phi)v^2+F(\zeta,\Phi,\zeta_g,\Phi_g)v v_g,
\end{eqnarray*}
where the F-functions are just collections of terms proportional to
different orders of $v$. When averaging over electron velocities, the
term $F(\zeta,\Phi,\zeta_g,\Phi_g)$ disappears since $\langle
v\rangle=0$ by definition. This is important, because the terms that
are left are just a sum of terms dependent on gas velocity
($v_g,\zeta_g,\Phi_g$) and terms dependent on thermal velocity
($v,\zeta,\Phi$). There are no cross terms dependent on both! So in
the radiative transfer equations for a moving electron gas, the part
dependent on temperature (thermal velocity) and the part dependent on
gas velocity are independent.

This simplifies the calculations since one can calculate the terms
dependent on thermal velocity and the terms dependent on gas velocity
separately. The terms containing the gas temperature can be found
assuming that the gas velocity is zero, and similarly the terms
dependent on gas velocity can be found assuming that the temperature
is zero. Then the different terms can be added to get the complete
radiative transfer equations for a moving electron gas with nonzero
temperature.

We now find the terms dependent on gas velocity assuming the
temperature to be zero. Going back to the Boltzmann collisional
equation in the rest frame of an electron [equations
(\ref{eq:boltzmann2})], one has to transform all quantities to the
frame of reference where the electron has a velocity $\mathbf{v_g}$.
This is exactly what was done previously, except that one has to
replace $(v,\zeta,\Phi)$ with $(v_g,\zeta_g,\Phi_g)$ in all the
expressions and expansions used, since the only velocity now is the
gas velocity.

Since there is no averaging over electron velocities and directions
this time (the electron gas has a constant velocity which all the
electrons are following), one can change to a coordinate system in
which the expressions and calculations are more easy. Previously, a
coordinate system $(x,y,z)$ with arbitrary axes was introduced. The
incoming and outgoing photon directions were described with polar and
azimuthal angles $(\eta=\cos{\theta},\phi)$ and
$(\eta'=\cos{\theta'},\phi')$ with respect to this coordinate system.
But since the directions of the axes were arbitrary, one can chose the
$z$-axis to be in the direction of the gas velocity. With this choice
of $z$-axis, the angle $\eta=\cos{\theta}$ between the $z$-axis and
the photon is just the angle $\zeta_g=\cos{\alpha_g}$ between the
photon and electron. The azimuthal angles are still called $\phi$ and
$\phi'$ for the photons with frequency $\nu$, and $\nu'$. With this
choice of axes, the change $\eta\rightarrow\zeta_g$ and
$\eta'\rightarrow\zeta_g'$ is made everywhere in the equations.

The correction factors $w$ and $u$ from the Lorentz transformations of
the angles $i_1$ and $i_2$ need special attention. These are expressed
(see appendix \ref{app:expansions}) through the angle $\Phi$. That was
a convenient angle when there was an integration over electron
directions. Now however, there is only an integration over incoming
photon directions (note that $d\Omega'=d\zeta_g'd\phi'$ with the
current choice of axes), so the angle $(\phi-\phi')$ is a better
choice than $\Phi$. The following relations between $\Phi$ and
$(\phi-\phi')$ can be found from geometrical considerations:
\begin{eqnarray*}
\cos{\Phi}&=&\frac{\zeta_g'-\zeta_g \mu}{\sqrt{1-\zeta_g'^2}\sqrt{1-\mu^2}},\\
\sin{\Phi}&=&-\frac{\sqrt{1-\zeta_g'^2}}{\sqrt{1-\mu^2}}\sin{(\phi-\phi')}.
\end{eqnarray*}

Since there is no integration over electron velocities, it is now easy
to find the equations describing the radiative transfer of polarized
radiation in a moving zero temperature electron gas. All that is
necessary is to put the expanded Lorentz transformed quantities
(appendix \ref{app:expansions}) into the Boltzmann equation
(\ref{eq:boltzmann2}), and replace $(\eta,\eta',\Phi)$ in the way
described above. The resulting equations are the radiative transfer
equations for a moving zero temperature electron gas. Adding the
temperature dependent terms from equation (\ref{eq:boltzmann3}) (with
a coordinate system where the $z$ axis is chosen to be along the gas
velocity direction) we find that
\begin{eqnarray}
&&\pdv{\n}{t}+\bomega\cdot\nabla\!\n=-\sigma_T\rho(1-v\zeta-2\nu)\n+\nonumber\\
&&\sigma_T\rho\frac{3}{16\pi}\int_{-1}^{1}d\zeta'\int_{0}^{2\pi}d\phi'\biggl\{\mt{L}_1\bigl([1+2\nu(1-\mu)]\mt{B}-2\tau\mt{C}\nonumber\\
&&+[\mt{G}+(2\zeta'-\zeta)\mt{B}]v+[\mt{H}+(2\zeta'-\zeta)\mt{G}-(1-3{\zeta'}^2\nonumber\\
&&+2\zeta\zeta')\mt{B}]v^2\bigr)\mt{L}_2\n'+\mt{L}_1\biggl([(\nu+4\tau)(1-\mu)+v(\zeta'-\zeta)]\mt{B}\nonumber\\
&&+(\zeta'-\zeta)[(3\zeta'-\zeta)\mt{B}+\mt{G}]v^2\biggr)\mt{L}_2\nu\frac{\partial\n'}{\partial\nu}+\frac{1}{2}\biggl\{v^2(\zeta'-\zeta)^2\nonumber\\
&&+2\tau(1-\mu)\biggr\}\nu^2\mt{L}_1\mt{B}\mt{L}_2\frac{\partial^2\n'}{\partial{\nu}^2}+2\mt{N}\mt{L}_1\mt{B}\mt{L}_2(1-\mu)\nu[2\n'\nonumber\\
&&+\nu\frac{\partial \n'}{\partial\nu}]\biggr\},
\label{eq:complete}
\end{eqnarray}
where we have written $\zeta$ instead of $\zeta_g$ for simplicity.
Here the matrices $\mt{G}$ and $\mt{H}$ come from the expansion of
$\mt{A}$, $\mt{A}=\mt{B}+\mt{G}v+\mt{H}v^2$. These matrices are given
in appendix \ref{app:matrices}. The $\mt{L}$ matrices are again
$\mt{L}_1=\mt{L}(\pi-i_2)$ and $\mt{L}_2=\mt{L}(-i_1)$, where the
angles $i_1$ and $i_2$ are given by equations (\ref{eq:zerothi12})
(remember to let $\eta\rightarrow\zeta_g$ and
$\eta'\rightarrow\zeta_g'$). This is the equations describing the
change of a radiation field upon interaction with a hot (first order
in $kT/mc^2$), moving (second order in $v_g/c$) electron gas to first
order in $h\nu/mc^2$. To be able to perform the integral over incoming
electron directions, one can use the relation
$\mu=\zeta_g\zeta_g'+\sqrt{1-\zeta_g^2}\sqrt{1-{\zeta_g'}^2}\cos{(\phi-\phi')}$.

\section{application to the SZ effect}
\label{sect:SZ}
One important application of the equations deduced here, is the
interaction of the cosmic background radiation with the hot electron
gas in clusters of galaxies. In this case, the frequency is so low
that only terms to zeroth order in frequency need to be taken into
account. For small changes in the radiation field, the initial field
can be inserted for $\n$ on the right hand side in equation
(\ref{eq:complete}) and integrated along a path through the cluster
\cite{sz1}. The initial radiation field is the cosmic background
radiation which can be assumed to be isotropic and homogenous. In this
case, the integration in equation (\ref{eq:complete}) over incoming
photons can easily be made using the techniques of Section
\ref{sect:symmetry}.

First we assume that the gas velocity is zero, studying only
temperature dependent effects. In this case the solution of equation
(\ref{eq:complete}) for small changes in the radiation field gives the
following change $\Delta \n$ to first order in optical depth:
\begin{eqnarray*}
\Delta \n&=&-s\n_0+\frac{3}{16\pi}\int ds\biggl(\int_-1^1d\zeta'\int_0^{2\pi}d\phi'\biggl\{\mt{L}_1\bigl[\mt{B}\\
&&-2\tau\mt{C}\bigr]\mt{L}_2\n_0+4\tau(1-\mu)\mt{L}_1\mt{B}\mt{L}_2\nu\pdv{\n_0}{\nu}\\
&&+\tau(1-\mu)\mt{L}_1\mt{B}\mt{L}_2\nu^2\pddv{\n_0}{\nu}\biggr\}\biggr),
\end{eqnarray*}
where $s=\sigma_T\rho$ is the optical depth and $\int ds$ is an
integral through the optical depth of the cluster. The initial
radiation field $\n_0$ is given by
\[
\n_0=\frac{1}{e^{\nu/\tau}-1}\left(
\begin{array}{c}
1\\0\\0\\0
\end{array}\right).
\] 
This is a matrix equation equivalent to four scalar equations each
describing the change in the Stokes parameters $I$,$Q$,$U$ and $V$.
Making the integrations, the first equation gives the thermal SZ
effect \cite{sz1}. The other equations all give zero polarization to
first order in optical depth as was excpected because of the isotropy
of the cluster gas and the radiation field.

The equation can also be used to study the effect of the intracluster
medium on radiation from extended radio sources within the cluster.
The state of polarization of the radiation coming from radio sources
will be changed upon scattering with electrons in the intracluster
gas. The change in the degree of polarization will of course be
proportional to the optical depth. It will also be dependent on the
temperature of the cluster gas, as can be seen by considering the
change of $\n$ from a single scattering [from equation
(\ref{eq:complete}) considering only the inscattering term]:
\begin{eqnarray*}
\Delta \n&=&\frac{3}{16\pi}\int_{-1}^1d\zeta'\int_0^{2\pi}d\phi'\biggl\{\mt{L}_1\bigl[\mt{B}-2\tau\mt{C}\bigr]\mt{L}_2\n_0\\
&&+4\tau(1-\mu)\mt{L}_1\mt{B}\mt{L}_2\nu\pdv{\n_0}{\nu}+\tau(1-\mu)\mt{L}_1\mt{B}\mt{L}_2\nu^2\pddv{\n_0}{\nu}\biggr\},
\end{eqnarray*}
where $\n_0$ now is the initial spectrum of radiation coming from the
radio source. This time $\n_0$ is not isotropic since the radiation is
only coming from some angles. A more detalied study is necessary to
determine if the change of polarization of radiation from a radio
source can be used to determine density and temperature of the cluster
gas.

Finally we consider the kinematic SZ effect which is the change of the
background radiation because of scattering on a moving cluster gas.
This time we assume that the temperature of the gas is zero. Again,
using equation (\ref{eq:complete}), we find to first order in optical
depth using an initial Planck function:
\begin{eqnarray*}
\frac{\Delta n}{n}&=&\frac{v}{c}\zeta_g\frac{xe^x}{e^x-1}s\\
\frac{\Delta n_Q}{n}&=&-\frac{1}{10}\frac{v^2}{c^2}(1-\zeta_g^2)s,
\end{eqnarray*}
where $x=\nu/\tau$. The first equation is just the formula for the
kinematic SZ effect which was given by Sunyaev \& Zeldovich
\shortcite{sz2} without derivation. The second equation shows that
some polarization is produced from the kinematic effect. The degree of
polarization is proportional to the tangential velocity component of
the cluster. This result was also given by Sunyaev \& Zeldovich
\shortcite{sz2}, again without deductions. The second result can be
used to determine the tangential velocity of clusters of galaxies, but
at the moment the polarization degree produced (about $10^{-8})$ is
too small to be observed.

\section{conclusion}
\label{sect:concl}
In equations (\ref{eq:boltzmann3}) we have obtained the radiative
transfer equations for the Stokes parameters to first order in
$h\nu/mc^2$ and $kT/mc^2$. Comparing these to those obtained by
Nagirner \shortcite {nagirner,nagirner2}, some small differences are
seen. Comparing the radiative transfer equations for an axisymmetric
field [equations (\ref{eq:I}) and (\ref{eq:Q})] with those obtained by
Stark \shortcite{stark}, we find several disagreements. However,
setting $Q=0$, studying only the intensity of radiation, the equations
actually agree. This indicates that the difference has to do with
polarization, and indeed, in the paper of Stark, we find that the
zeroth order rotation angles $i_1$ and $i_2$ were used instead of
$i_{1e}$ and $i_{2e}$. The relativistic corrections to these angles
were neglected, which, as is seen from the Taylor expansion, is
altering the integration substantially. Including these corrections,
the equations in the paper of Stark become identical to those obtained
here [equations (\ref{eq:I}) and (\ref{eq:Q})], except for the induced
scattering terms. By carefully performing the last integration in that
paper, we find that the induced scattering terms also agree with what
we have derived here.

Manipulating the equations in the paper of Nagirner
\shortcite{nagirner} before integration, it is possible to show that
they agree with equation (\ref{eq:boltzmann2}) of this paper. So there
has to be an error somewhere in his integration. And indeed, by
correcting the misprints in the integration, we find full agreement
with the results presented here. So our equations
(\ref{eq:boltzmann3}) correct the errors of Stark \shortcite{stark}
and Nagirner \shortcite{nagirner,nagirner2}.

We also transformed the equations to a frame where the electron gas is
moving [equations (\ref{eq:complete})]. In this frame the equations
were derived to second order in gas velocity. The transfer equations
were tested by showing that they give zero polarization produced from
the thermal SZ effect to first order in optical depth. We also showed
that polarization from the kinematic SZ effect is very small.The
formalism developed here can probably be used to study polarization of
radiation from radio sources in the cluster upon scattering with the
intracluster medium.

\section*{acknowledgements}
We thank Mark Birkinshaw for suggesting the study of polarization in
the SZ-effect and Juri Poutanen for help and comments. We also thank
the Research Council of Norway for a travel grant.

\begin{appendix}

\section{Taylor expansions}
\label{app:expansions}
The following quantities expanded to first order in $\nu$ and second order in $v$ are needed in Section \ref{sect:integration}:
\begin{eqnarray}
\label{eq:expanded}
\left(\frac{\nu_e''}{\nu_e}\right)^2&\approx&1-2\nu(1-\mu),\nonumber\\
\left(\frac{d\nu_e'}{d\nu_e}\right)&\approx&1+2 \nu(1-\mu),\nonumber\\
\frac{(\nu'-\nu)}{\nu}&\approx&\nu(1-\mu)+v(\zeta'-\zeta)+v^2\zeta'(\zeta'-\zeta),\nonumber\\
\frac{(\nu''-\nu)}{\nu}&\approx&-\nu(1-\mu)+v(\zeta'-\zeta)+v^2\zeta'(\zeta'-\zeta),\nonumber\\
\frac{(\nu'-\nu)^2}{\nu^2}&\approx&\frac{(\nu''-\nu)^2}{\nu^2}\approx v^2 (\zeta'-\zeta)^2,\nonumber\\
\frac{d\Omega_e'}{d\Omega'}&\approx&1+2 v \zeta'-v^2(1-3 \zeta'^2),\nonumber\\
\gamma&\approx&1+\frac{1}{2}v^2,\nonumber\\
\cos{\Theta_e}&\approx&\mu-(1-\mu)(v(\zeta+\zeta')\nonumber\\&&+v^2(\zeta^2+\zeta'^2+\zeta\zeta'-1)),\nonumber\\
\cos^2{\Theta_e}&\approx&\mu^2-v 2(1-\mu)\mu(\zeta+\zeta')\nonumber\\&&-v^2 (1-\mu)(-2\mu-\zeta^2+3\mu\zeta^2\nonumber\\&&-2\zeta\zeta'+4\mu\zeta\zeta'-\zeta'^2+3\mu\zeta'^2),\nonumber\\
w&\approx&1+2v^2\frac{1-\mu}{1+\mu}(\zeta^2-1)(\cos^2{\Phi}-1)\nonumber\\
u&\approx&2v(1-\mu)\sqrt{\frac{1-\zeta^2}{1-\mu^2}}\sin{\Phi}+\nonumber\\
&&2v^2\frac{1-\mu}{1+\mu}\sqrt{\frac{1-\zeta^2}{1-\mu^2}}(\zeta+\zeta')\sin{\Phi},
\end{eqnarray}
where $u\equiv 2xy$ and $w\equiv (2x^2-1)$.

\section{matrices for the radiative transfer equations}
\label{app:matrices}
The following matrices were used in equation (\ref{eq:boltzmann3}) and (\ref{eq:complete}):
\[
\mt{B}=\left(
\begin{array}{cccc}
\mu^2+1&\mu^2-1&0&0\\
\mu^2-1&\mu^2+1&0&0\\
0&0&2\mu&0\\
0&0&0&2\mu\\
\end{array}
\right),
\]
\[
\mt{C}=\left(
\begin{array}{cccc}
C_{11}&C_{12}&0&0\\
C_{12}&C_{22}&0&0\\
0&0&C_{33}&0\\
0&0&0&C_{44}\\
\end{array}
\right),
\]
\begin{eqnarray*}
C_{11}=-1+2\mu+3\mu^2-2\mu^3,&C_{12}=-2(\mu-1)^2(\mu+1),\\
C_{22}=1+2\mu+\mu^2-2\mu^3,&C_{33}=2+2\mu-2\mu^2,\\
C_{44}=2\mu.&\\
\end{eqnarray*}
\[
\mt{G}=\left(
\begin{array}{cccc}
G_{11}&G_{11}&G_{13}&0\\
G_{11}&G_{11}&G_{23}^0\\
-G_{13}&-G_{23}&G_{33}&0\\
0&0&0&G_{33}
\end{array}
\right),
\]
\[
\mt{H}=\left(
\begin{array}{cccc}
H_{11}&H_{12}&H_{13}&0\\
H_{12}&H_{22}&H_{23}^0\\
-H_{13}&-H_{23}&H_{33}&0\\
0&0&0&H_{44},
\end{array}
\right),
\]
\begin{eqnarray*}
G_{11}&=&2(\mu-1)\mu(\zeta_g+\zeta_g'),\\
G_{13}&=&2(\mu-1)\sqrt{1-\zeta_g^2}\sqrt{1-{\zeta_g'}^2}\sin{(\phi-\phi')},\\
G_{23}&=&2(\mu+1)\sqrt{1-\zeta_g^2}\sqrt{1-{\zeta_g'}^2}\sin{(\phi-\phi')},\ \ \ G_{33}=\frac{G_{11}}{\mu},\\
H_{11}&=&(1-\mu)[(\zeta_g+\zeta_g')^2-\mu(-2+3\zeta_g^2+4\zeta_g\zeta_g'+3{\zeta_g'}^2)],\\
H_{12}&=&\frac{\mu-1}{\mu+1}[-2+(\zeta_g-\zeta_g')^2+2\mu(-1+\zeta_g^2-\zeta_g\zeta_g'+{\zeta_g'}^2)\\
&&+\mu^2(3\zeta_g^2+4\zeta_g\zeta_g'+3{\zeta_g'}^2)],\\
H_{22}&=&-4+2\mu+2\mu^2+(5-4\mu+3\mu^2)(\zeta_g^2+{\zeta_g'}^2)\\
&&+(2-14\mu+4\mu^2)\zeta_g\zeta_g',\\
H_{33}&=&-2-2\mu+4\mu^2+2(1+\mu)(\zeta_g^2+{\zeta_g'}^2)-2(1+3\mu)\zeta_g\zeta_g',\\
H_{13}&=&\frac{1+2\mu}{1+\mu}(\zeta_g+\zeta_g')G_{13},\ \ \ H_{23}=\frac{(2\mu-1)}{\mu-1}(\zeta_g+\zeta_g')G_{23},\\
H_{44}&=&2(1-\mu)(1-\zeta_g^2-{\zeta_g'}^2-\zeta_g\zeta_g').
\end{eqnarray*}

\end{appendix}

\label{lastpage}
\end{document}